\documentclass[fleqn,usenatbib]{mnras}

\usepackage{newtxtext,newtxmath}
\usepackage[T1]{fontenc}

\DeclareRobustCommand{\VAN}[3]{#2}
\let\VANthebibliography\thebibliography
\def\thebibliography{\DeclareRobustCommand{\VAN}[3]{##3}\VANthebibliography}

\usepackage{graphicx}	% Including figure files
\usepackage{amsmath}	% Advanced maths commands
\usepackage{subfigure}
\usepackage{multirow}
\newcommand\be{\begin{equation}}
\newcommand\e{\end{equation}}
\newcommand\ba{\begin{eqnarray}}
\newcommand\ay{\end{eqnarray}}
\newcommand\nn{\nonumber}

\newcommand\cab{\text{C}_{2_1}^0}
\newcommand\cac{\text{C}_{2_2}^0}
\newcommand\cad{\text{C}_{2_3}^0}

\newcommand\ccb{\text{C}_{2_1}^2}
\newcommand\ccc{\text{C}_{2_2}^2}
\newcommand\ccd{\text{C}_{2_3}^2}

\newcommand\pd{\partial}
\newcommand\eps{\epsilon}

\newcommand\pss{|\psi|^2}

\title[2D modulated ESWs in non-Maxwellian dusty plasmas]{On the stability of two-dimensional modulated electrostatic wavepackets in non-Maxwellian dusty plasma -- application in Saturn's magnetosphere}

\author[K. Singh et al.]{
Kuldeep Singh,$^{1}$\thanks{E-mail: singh.kdeep07@gmail.com; kuldeep.singh@ku.ac.ae (KS)}
Michael McKerr,$^{1,2,3}$
and Ioannis Kourakis$^{1,4}$
\\
$^{1}$Department  of  Mathematics,  Khalifa  University  of  Science  \&  Technology,  Abu  Dhabi,  UAE.\\
$^{2}$Department of Sciences and Engineering, Sorbonne University Abu Dhabi, UAE.\\
$^{3}$14 Portna Road, Kilrea, County Derry, BT51 5SW, Northern Ireland, UK.\\
$^{4}$Space and Planetary Science Center, Khalifa University, Abu Dhabi, UAE.\\
}

\date{Accepted XXX. Received YYY; in original form ZZZ}

\pubyear{2015}

\begin{document}
\label{firstpage}
\pagerange{\pageref{firstpage}--\pageref{lastpage}}
\maketitle

\begin{abstract}
Motivated by observations of localized electrostatic wavepackets by the Voyager 1 and 2 and Cassini missions in Saturn's magnetosphere, we have investigated the evolution of modulated electrostatic wavepackets in a dusty plasma environment. The well known dust-ion acoustic (DIA) mode was selected to explore the dynamics of multi-dimensional structures, by means of a Davey–Stewartson (DS) model, by taking into account the presence of a highly energetic (suprathermal, kappa-distributed) electron population in combination with heavy (immobile) dust in the background. The modulational (in)stability profile of DIA wavepackets for both negative as well as positive dust charge is investigated. A set of explicit criteria for modulational instability (MI) to occur is obtained. Wavepacket modulation properties in 3D dusty plasmas are shown to differ from e.g. Maxwellian plasmas in 1D. Stronger negative dust concentration results in a narrower instability window in the $K$ (perturbation wavenumber) domain and to a suppressed growth rate. In the opposite manner, the instability growth rate increases for higher positive dust concentration and the instability window gets larger. In a nutshell, negative dust seems to suppress instability while positive dust appears to favor the amplitude modulation instability mechanism. Finally, stronger deviation from the Maxwell-Boltzmann equilibrium, i.e. smaller $\kappa_e$ values, lead(s) to stronger instability growth in a wider wavenumber window -- hence suprathermal electrons favor MI regardless of the dust charge sign (i.e. for either positive or negative dust). The  wavepacket modulation properties in 2D dusty plasmas thus differ from e.g. Maxwellian plasmas in 1D, both quantitatively and qualitatively, as indicated by a generalized dispersion relation explicitly derived in this paper (for the amplitude perturbation). Our results can be compared against existing experimental data in space, especially in Saturn's magnetosphere.
\end{abstract}


%%%%%%%%%%%%%%%%%%%%%%%%%%%%%%%%%%%%%%%%%%%%%%%%%%

%%%%%%%%%%%%%%%%% BODY OF PAPER %%%%%%%%%%%%%%%%%%

\section{Introduction}

Dust is an ineluctable ingredient in space and astrophysical environments. The last few decades have seen a growing interest in elucidating the physics of dusty plasma and in investigating the associated (e.g. electrostatic) modes and instabilities, because of their essential role in space and astrophysical plasmas (e.g., in planetary rings, interior of heavy planets, etc.) \citep{goe89,hor86,ver96} and in laboratory plasmas (e.g., fusion devices, plasma devices, solar cells, semiconductor chips etc.) \citep{sam01,sam05,adh07}. A dusty plasma consisting of normal (electrons-ions) plasma in addition to charged (positive or negative) dust particles. \citet{shukla92} first  time theorized the existence of dust ion acoustic waves which were further verified experimentally by \citet{bar96}. The electron thermal pressure provides restoring force and the inertia by ion mass.The phase speed of DIAWs is larger than the usual ion acoustic speed because $n_{i0} > n_{e0}$ for negative dust. A number of investigations to study the propagation properties of dust ion acoustic nonlinear coherent structures in different plasma environments have been reported \citep{mam08,mam09,ali11,saini13}.

\par On the other hand, satellite observations have confirmed the ubiquitous presence of energetic particle populations e.g. in the solar wind, manifested in long tailed (non-Maxwellian) distributions. However, satellite missions have indicated that there are many regions in space plasmas where charged species deviate from Maxwellian behavior \citep{liu09}. These suprathermal species have been found in the solar wind, magnetosphere \citep{fel75}, interstellar medium, auroral zone plasma \citep{laz08,men94}, and also in the terrestrial magnetosheath \citep{mas06,qur19}. Vasyliunas used the kappa velocity distribution for the first time as an empirical formula to fit data from the spacecraft OGO 1 and OGO 3 in the terrestrial magnetosphere \citep{vas68} presenting a long-tailed behavior in the superthermal component of the distribution. Since then, it has been used to fit data from spacecraft in Saturn, solar wind \citep{arm83}, Earth's magnetospheric plasma and Jupiter \citep{leu82}. Space observations usually adopt\ a kappa distribution with low values of $\kappa$ (between 2 to 6 in Saturn)\citep{sch08}. For larger spectral index values (i.e., $\kappa\rightarrow\infty$), the non-Maxwellian kappa distribution reduces to the Maxwellian distribution. The Voyager 1 and 2 spacecraft obtained data from Saturn's magnetosphere reveals that ions follow power law at high energies. \citet{kri83} used kappa distributions to fit data observations for ions in the Saturn magnetosphere, with spectral index values ranging from 6 to 8. In addition, the Cassini team collected data from spacecraft orbiting Saturn and covering distances ranging from 5.4-18 $R_s$, where $R_s$ is the radius of Saturn ($R_s \thickapprox 60268km$). The observed data is well fitted by electrons with kappa distribution in Saturn's magnetosphere \citep{sch08}.

The Radio and Plasma Wave Science (RPWS) instrument onboard Cassini has provided strong corroboration that charged dust grains in the E-ring interacts collectively with the surrounding plasma of Saturn's magnetosphere \citep{wahlund09}. Based on data from grain sizes 41 mm, it has been deduced that Enceladus ejects a dust torus that disperses to form most of the E-ring. Dust densities of these larger grains are of the order of $10^{-1}$ m$^{-3}$, which is small compared to plasma densities in the surrounding plasma disk. However, the dust population follows a $r_d^{-m}$ power law (with $m \sim 4-5$) \citep{kempf05}, where $r_d$ is the dust grain radius, and dust densities therefore rise sharply for smaller dust particles, hence the E-ring is by far dominated by sub-millimeter-sized grains. Furthermore, the charge of the dust in the E-ring has been measured by the dust experiment (CDA) to be a few volts negative inside 7 $R_s$ \citep{kempf06}. Electrostatic coupling of micron-sized dust has also been inferred in the more visible inner rings of Saturn in the form of ``spokes" \citep{wahlund09}. Cassini Radio and Plasma Wave Science Wideband Receiver (WBR) data specifically observed high percentage of bipolar-type electrostatic solitary waves (ESWs)  in the range of less than 10 $R_s$ in the years 2004-2008. This location is consistent with the densest part of Saturn's E ring and Enceladus's orbit\citep{pic15}. Typical plasma parameters corresponding to planetary rings for dusty plasma are:
$n_{i0}=  (10^{13}- 10^{16}) m^{-3}$, $n_{e0} = (10^{13}- 10^{16}) m^{-3}$, $Z_d = 10^3$ , $n_d = 10^{13} m^{-3}$ , $T_e = 50$ eV, $T_i = 0.05$ eV \citep{goe89}.

Modulated wavepackets are ubiquitous in space; however, the generalization of the standard theory for modulational processes \citep{ik05} to higher dimensionality \citep{ds,Nishinari,Nishinari94,Fokas,aDuan,bDuan,Ghosh,xue04,yashika} is not entirely understood.  The transverse perturbations are supported by higher dimensions. Introducing  transverse perturbation results in the generation of an anisotropy in the system, which  impacts on wave propagation.This fact motivated our study of modulated dust-ion acoustic (DIA) wavepackets in non-Maxwellian dusty plasmas in a 2D or 3D geometry. A representative evolution equation for modulated wavepackets in space  is the so-called Davey Stewartson system (DS) \citep{ds}, which is a two-dimensional generalization of the nonlinear Schr\"{o}dinger equation (NLSE). While the NLSE finds extensive applications in plasma physics, including the modelling of several wave phenomena in space plasmas, the DS equations have received scant attention from plasma and space scientists. A rare exception is a series of papers by  \citet{Nishinari,Nishinari94}, who have shown, with the help of a suitable reductive perturbation method, that the nonlinear evolution of an ion acoustic wave 2D wavepacket propagating in non-magnetized plasma modeled by a DS-II system \citep{Nishinari}, while the same formulism applied  to magnetized plasma leads to a DS-I system that possesses dromions solution \citep{Nishinari94}. We follow here the classification by  \citet{Fokas} who showed that four types of DS systems exist in general. Similar considerations might, therefore, apply for other modes in a plasma, particularly when their nonlinear stationary states display two- or three-dimensional structures. Our present
work is motivated by such considerations particularly in the context of space plasma observations which provide a rich
source of such potential structures.  \citet{bDuan} analyzed higher order transverse perturbations for modulated wave packets in a dusty plasma by deriving Davey–Stewartson equation. The instability for small amplitude linear transverse perturbations has also been examined.  \citet{Ghosh} examined the coupled Davey-Stewartson I equations for electron acoustic waves in the context of PCBL region which admit exponentially localized solutions called dromions. \citet{xue04} investigated the  modulation of dust ion acoustic wave (DIAW) in unmagnetized dusty plasmas with the transverse plane and derive a three-dimensional Davey–Stewartson (3D DS) equation. It may be highlighted that modulation properties of DIAW in 3D dusty plasmas are very different from that in 1D case. Relying on a multiple scale perturbation technique, a two-dimensional Davey–Stewartson (DS) equation is obtained, for the evolution of modulated electrostatic wavepackets in plasmas, taking into account the presence of a superthermal (kappa-distributed) electron background. The modulational (in)stability profile of DIA wavepackets is investigated. A set of explicit criteria for modulational instability to occur is obtained. Wavepacket modulation properties in 3D dusty plasmas are shown to differ from e.g. Maxwellian plasmas in 1D. Our results can be compared against existing experimental data in space (and hopefully motivate new ones), especially in Saturn's magnetosphere and in cometary tails \citep{goe89}.
\section{Fluid model}

We consider an unmagnetized collisionless plasma comprising of inertial ions, non-Maxwellian electrons  and immobile dust. The fluid model equations include the \\
continuity equation: \be \frac{\pd N_{i}}{\pd t'}  +\nabla'\cdot(N_{i}\vec V_{i})=0,\label{1}\e
the momentum  equation:\be \frac{\pd\vec V_{i}}{\pd t'}+\vec V_{i}\cdot \nabla' \vec V_{i}=-\frac{Z_{i}e}{m}\nabla' \Phi\label{2}\e
and Poisson's equation:\be \nabla'^2\Phi=-4\pi e(Z_{i}N_{i}-Z_{e}N_{e}+s_{d}Z_{d0}N_{d0}).\label{3}\e
The electron density is written by:
\be N_{e}=n_{e0}\left(1-\frac{e \Phi}{k_B T_e(\kappa_{e}-\frac{3}{2})}\right)^{-\kappa_{e}+\frac{1}{2}}\label{4}.\e
 The charge neutrality condition at equilibrium imposes
\be n_{e0} - s_{d}Z_{d0} n_{d0} = Z_i n_{i0}  \label{QN} \e
where $n_{\jmath0}$ for ($\jmath=i,e,d$) are unperturbed number density for electron,  positron,  ion and dust respectively and $ Z_{d0}$ is the number of negative elementary charge on dust grains. Here, $s_{d}=+1$ for positively charged dust and $s_d=-1$ for negatively charged dust.\\

In order to make further analysis easier, Eqs. (\ref{1})-(\ref{4}) are re-scaled by introducing the following dimensionless variables: the number density $n_j = \frac{N_{j}}{n_{i0}}(j=i,e,d)$; velocity $v_{i} = \frac{V_{i}}{C_{i}}$ (i.e., $C_{i}=(Z_{i}k_{B}T_{e}/m_{i})^{1/2}$); time $t = t' \omega_{pi}$(i.e., $\omega_{pi}=(4\pi e^{2}Z_{i}^{2}n_{i0}/m_{i})^{1/2}$ ); space derivative operator (divergence vector) $\nabla = {\lambda_{Di}} {\nabla'} $ (i.e., $\lambda_{Di}=(k_{B}T_{e}/4\pi e^{2}Z_{i}n_{i0})^{1/2}$ ); electrostatic potential $\phi =  \frac{e\Phi}{k_B T_e}$. Therefore, the charge neutrality condition can be written as:
\[\delta_e =\frac{n_{e0}}{Z_{i}n_{i0}}= 1+ s_d \delta_d \, , \]
where $\delta_{d}=\frac{Z_{d0}n_{d0}}{Z_{i}n_{i0}}$. An \emph{e-i} (i.e., dust free) plasma is recovered for $\delta_d=0$. \\
In the following, all quantities will be dimensionless, unless otherwise stated. The fluid model Eqs. (\ref{1})-(\ref{4}) become
\begin{eqnarray} \frac{\pd n_{i}}{\pd t}  +\nabla\cdot(n_{i}\vec v_{i})=0,\label{5} \\
\frac{\pd\vec v_{i}}{\pd t}+\vec v_{i}\cdot \nabla \vec v_{i}=-\nabla\phi, \label{6} \\
\nabla^2\phi=n_{e}- s_{d}\delta_d-n_{i},\label{7} \end{eqnarray}
where the normalized expression for electron density is
\be n_{e}=\delta_e\left(1-\frac{\phi}{\kappa_{e}-\frac{3}{2}}\right)^{-\kappa_{e}+\frac{1}{2}}\approx \delta_{e}+c_1\phi +c_2\phi^2+c_3\phi^3.\label{4a}\e
Here, $c_1=\delta_e\frac{\kappa_e-\frac{1}{2}}{(\kappa_e-\frac{3}{2})}$, \ $c_2=\delta_e\frac{\kappa_{e}^{2}-\frac{1}{4}}{2(\kappa_{e}-\frac{3}{2})^{2}}$, \ $c_3=\delta_e\frac{(\kappa_{e}^{2}-\frac{1}{4})(\kappa_{e}+\frac{3}{2})}{6(\kappa_{e}-\frac{3}{2})^{3}}$. Note that $c_{1,2,3}>0$ for all values of $\kappa_e$ and $\delta_e$ (or $\delta_d$).
Therefore, near equilibrium, Poisson's Eq. (\ref{7}) becomes
\be \nabla^2\phi \simeq (1-n_{i})+c_1\phi +c_2\phi^2+c_3\phi^3 \, , \label{7a}\e
where all coefficient were defined above. The quasi-neutrality condition (\ref{QN}) (valid at equilibrium) was used to simplify the latter equation.

\section{Perturbative analysis}

We proceed by expanding the state variables around equilibrium as
\ba \phi&=&\eps\phi_1+\eps^2\phi_2+\eps^3\phi_3+...\nn\\
v_i&=& \eps v_1+\eps^2 v_2+\eps^3 v_3+...\nn\\
n_i&=& 1+\eps n_1+\eps^2n_2+\eps^3 n_3+...\ay
and by introducing multiple evolution scales considered for the independent (time, space) variables as $ T_k =\eps^k t$ and   $\vec{X_k}=\eps^k \vec{x}$ where $k=0,1,2,3...$.
At every order $\eps^k$, the state variables are expanded as
\ba \phi_k&=& \sum_{m=-k}^{k}\phi_{k}^{(m)}e^{\iota m (\vec k \vec X_0 - \omega T_0)}  \nn\\
v_k&=& \sum_{m=-k}^{k}v_{k}^{(m)}e^{\iota m(\vec k \vec X_0 - \omega T_0)} \nn \\
n_k&=& \sum_{m=-k}^{k}n_{k}^{(m)}e^{\iota m(\vec k \vec X_0 - \omega T_0)} \, , \ay
where the phase obviously depends on the zeroth-order (fast) variables, while the harmonic amplitudes are assumed to depend only on the slower scales (for $m = 1, 2, ...$).

In order   $\eps$, we obtain the dispersion relation:
\be \omega^2=\frac{k^2}{c_1+k^2} \, , \label{dispersion}\e
where $k^2=k_{x}^{2}+k_{y}^{2}$. The leading (first) order harmonic amplitudes are expressed as $v_{1}^{(1)}= \frac{\vec k}{\omega}\psi$ and $n_{1}^{(1)}=\frac{k^2}{\omega^2}\psi$ in terms of the electrostatic potential disturbance, say, $\psi=\phi_{1}^{(1)}$.
\begin{figure*}
\centering
\subfigure[]{\includegraphics[width=2.5in]{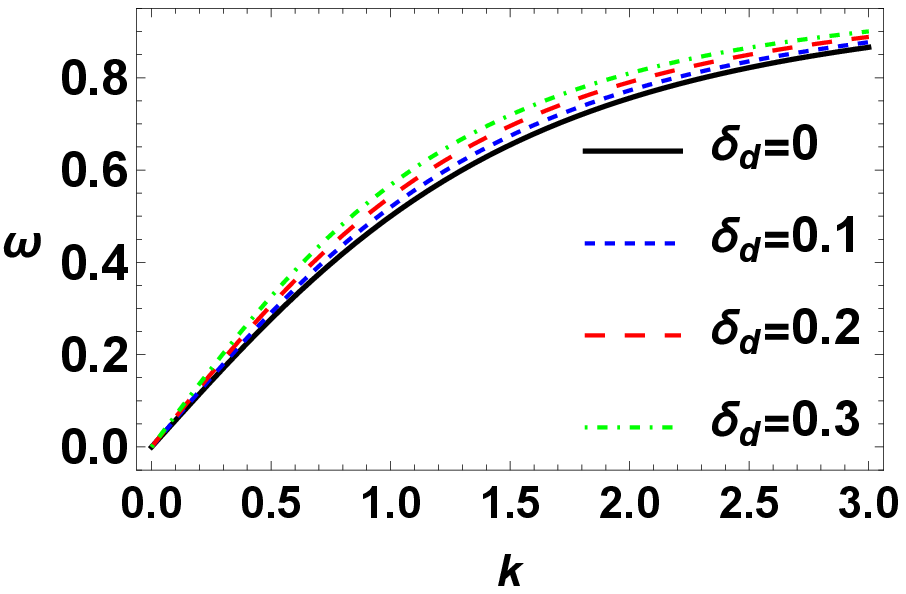}}
\subfigure[]{\includegraphics[width=2.5in]{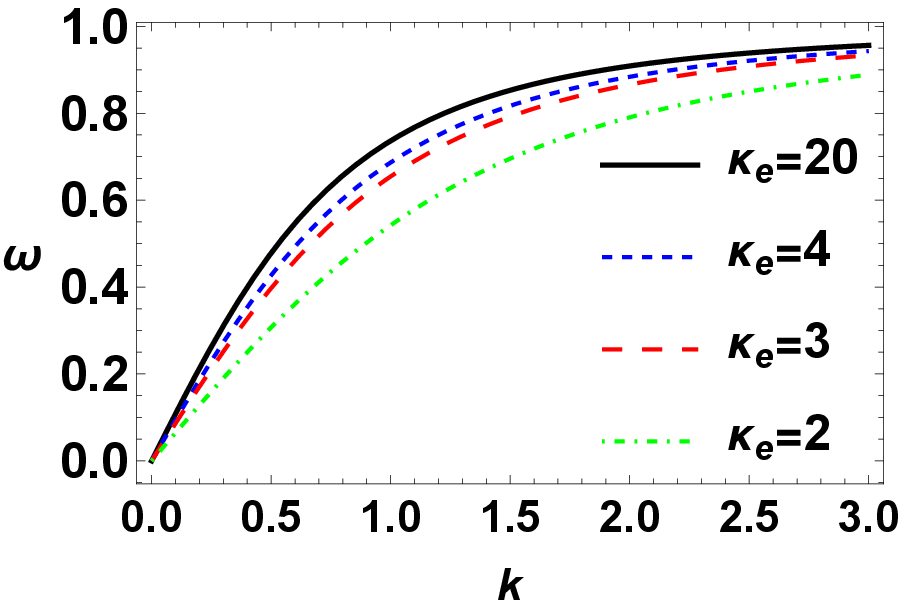}}
%\large{(a)} \hspace{3in} \large{(b)}
\caption{Plot of $\omega$ vs. $k$ for different values of (a) $\delta_d$  for fixed value of $\kappa_e=2$, (b) $\kappa_e$ for fixed value of $\delta_d=0.2$.}\label{f1}
\end{figure*}

\begin{figure*}
\centering
\subfigure[]{\includegraphics[width=2in,height=1.8in]{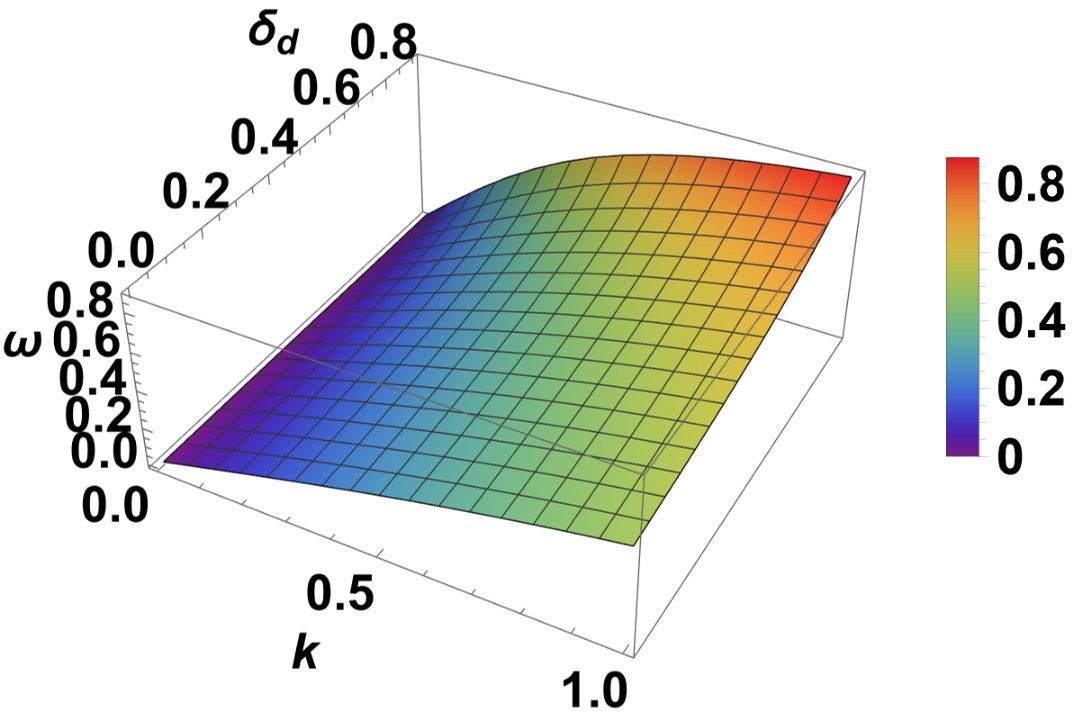}}
\subfigure[]{\includegraphics[width=2in,height=1.8in]{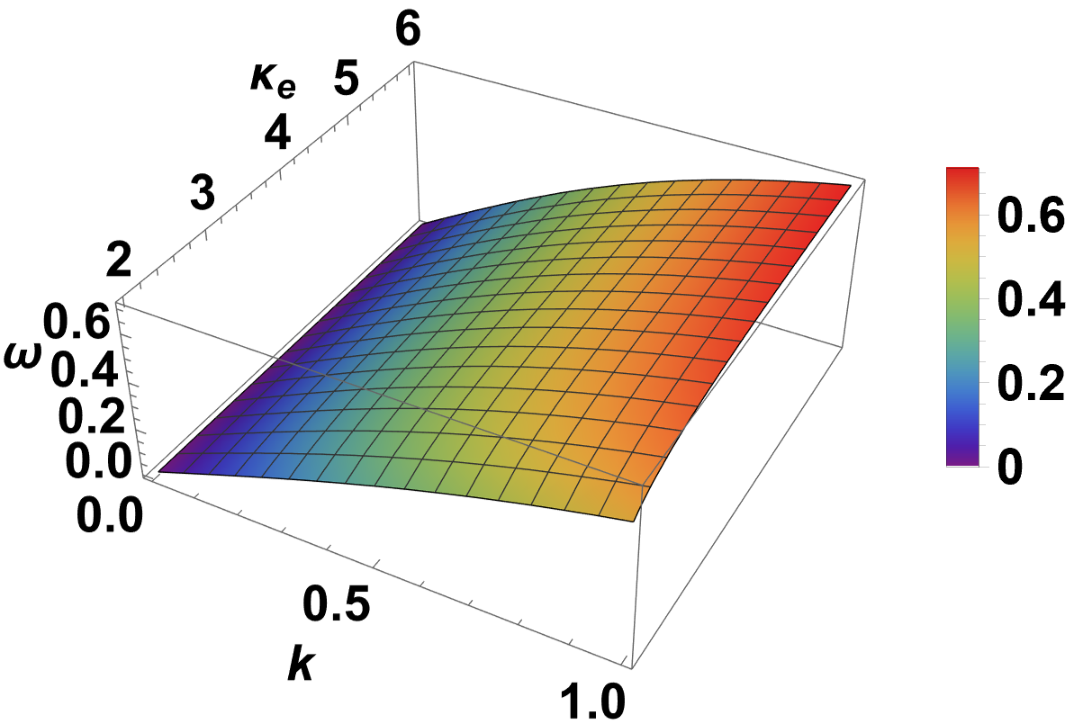}}
\subfigure[]{\includegraphics[width=2in,height=1.8in]{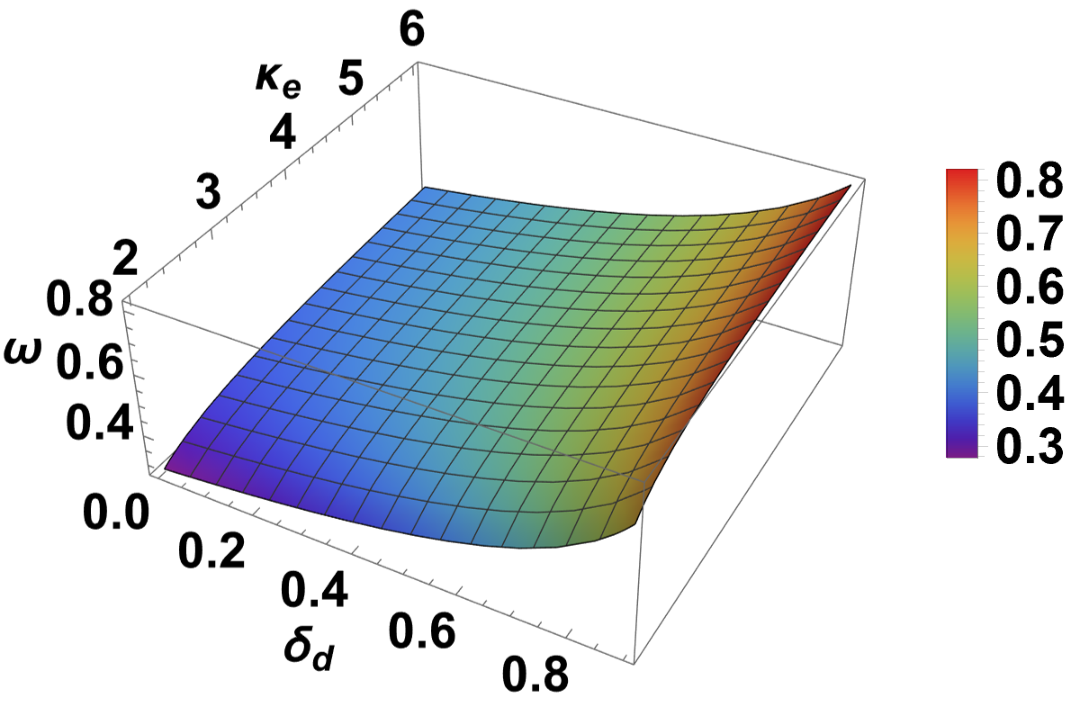}}
\caption{3D Plot of $\omega$ in the (a) $\delta_d$-$k$ plane for $\kappa_e=2$; (b) $\kappa_e$-$k$ plane for $\delta_d=0.2$; (c) $\kappa_e$-$\delta_d$ plane for $k=0.5$.}\label{f1b}
\end{figure*}

In order $\eps^2$, the requirement for suppression of secular terms leads to a condition in the form:
% $$\text{det}\left(\begin{array}{cc}-\omega^2& 2i\frac{k^2}{\omega}\frac{\pd}{\pd T_1}\psi+2i\vec k\cdot\nabla_1\psi \\ 1 & -2i\vec k\cdot\nabla_1\psi\end{array}\right)=0$$
\be \frac{\pd \psi}{\pd T_1}+ \frac{c_1\omega^3}{k^4}\vec k \cdot \nabla_1 \psi=0\label{kramer} \, . \e
The group velocity is thus prescribed as:
\be \vec v_g=\nabla_{\vec k}\omega=\frac{c_1\omega^3}{k^4}\vec k \, \label{vg}\e
In other words, the algebraic analysis provides the conclusion that the envelope moves at the group velocity, as expected.
This algebraic requirement suggests that the amplitude(s) of all state variable harmonics at this order will depend on a moving space coordinate (only), namely $X_1-\vec v_g T_1$, physically reflecting the fact that the amplitude (envelope) moves at the group velocity to leading nonlinear order ($\epsilon^2$), viz.
$\psi=\psi(\vec X_1-\vec v_g T_1; \vec X_2,  T_2)$ for the electrostatic potential (and analogous expressions for all other amplitudes).
This variable transformation, in the context of our multiple scale perturbation method, has been adopted (and its physical meaning has been discussed) in a number of monographs or articles  in plasma dynamics -- see e.g. \citet{InfeldRowlands} or \citet{ik05} for a space modeling context -- and also in nonlinear optics; see e.g. \citet{NewellBook}.

Solving the equations arising in this order, a set of expressions for the 2$^{nd}$ order zeroth, 1$^{st}$ and  2$^{nd}$ harmonics are obtained. It is convenient to express the density and fluid speed amplitudes in terms of the electrostatic potential amplitude, for all harmonic orders.
Setting $\phi_2^{(1)}=0$ with no loss of generality (since the first harmonic amplitude is left arbitrary in the algebra), the respective first harmonic amplitudes are given by:
\ba n_2^{(1)}&=&-2i\vec k\cdot \nabla_1\psi\nn\\
\vec v_2^{(1)}&=&\frac{i\omega c_1}{k^4}\vec k \vec k\cdot \nabla_1\psi-\frac{i}{\omega}\nabla_1\psi\ay

From the  2$^{nd}$ order 2$^{nd}$ harmonics, we obtained the respective second harmonic amplitudes as:

\ba \phi_2^{(2)}&=& \left(\frac{k^2}{2\omega^4}-\frac{c_2}{3k^2}\right)\psi^2=\ccd\psi^2\nn\\
n_2^{(2)}&=&\left(\frac{k^2}{\omega^2}\ccd+\frac{3k^4}{2\omega^4}\right)\psi^2=\ccb\psi^2\nn\\
\vec v_2^{(2)}&=& \left(\frac{\ccd}{\omega}+\frac{k^2}{2\omega^3}\right)\psi^2\vec k=\ccc\psi^2\vec k\label{ccd}\ay

The zeroth harmonic amplitudes (to second order) are not conclusively determined this order, so one needs to resort to  the third order equations ($\eps^3$) to find their analytical expression.
In order to find a relation between zeroth harmonic terms, we have chosen coordinate axes such that
\be \vec k=\left(\begin{array}{c} k\\0\end{array}\right),\e
i.e., considering propagation along the x-axis.
We express $\vec v_2^{(0)}$ in simpler terms for clarity as:
\be \vec v_2^{(0)}=\left(\begin{array}{c}u\\w\end{array}\right) \, .\e

The expanded fluid equations at zeroth order can then be solved in terms of $\pss$ and $Y=\int dx\frac{\pd w}{\pd y}$ to find:
\ba\phi_2^{(0)}&=& \frac{1}{c_1v_g^2-1}\left(\frac{2v_gk^3}{\omega^3}-2c_2v_g^2+\frac{k^2}{\omega^2}\right)\pss+\frac{v_g}{c_1v_g^2-1}Y \nn\\ &=&\cad\pss+\gamma_\phi Y\nn\\
n_2^{(0)}&=& \left(c_1\cad+2c_2\right)\pss+c_1\gamma_\phi Y=\cab\pss+\gamma_nY\nn\\
u_2^{(0)}&=&\frac{1}{v_g}\left(\cad+\frac{k^2}{\omega^2}\right)\pss+\frac{\gamma_\phi}{v_g}Y=\cac\pss+\gamma_uY\ay
with \be\begin{array}{cc} \cad=\frac{1}{c_1v_g^2-1}\left((3c_1+k^2)-2c_2v_g^2\right); &\gamma_\phi=\frac{v_g}{c_1v_g^2-1}\\
\cab=\left(c_1\cad+2c_2\right); &\gamma_n=c_1\gamma_\phi\\
\cac=\frac{1}{v_g}\left(\cad+\frac{k^2}{\omega^2}\right); &\gamma_u=\frac{1}{v_g}\gamma_\phi\end{array}\label{cad}\e
The integration constants in the above expressions are zero. If this is not the case, then a term proportional to $\psi$ appears in the first DS equation, which can be removed by a phase shift on $\psi$.

\begin{figure*}
\centering
\subfigure[]{\includegraphics[width=2.5in]{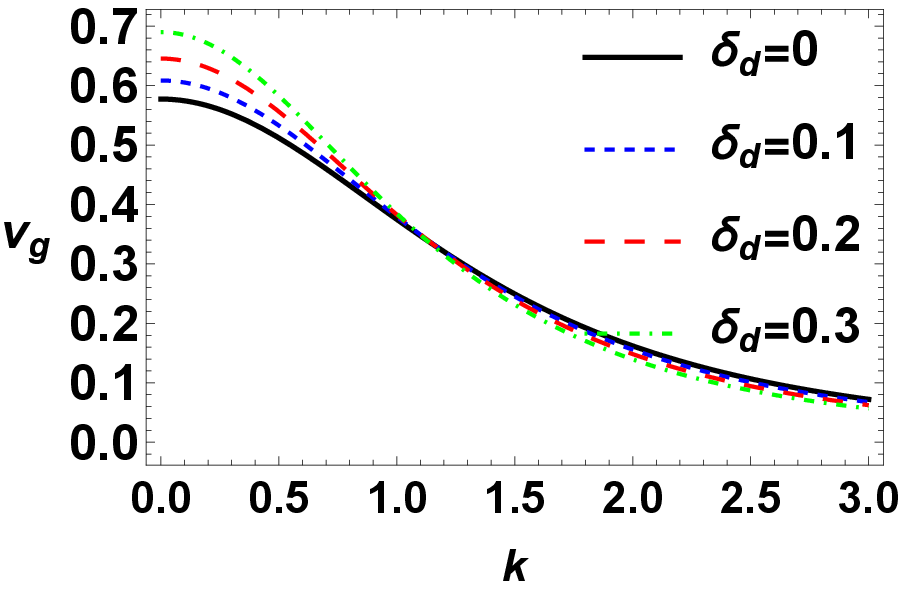}}
\subfigure[]{\includegraphics[width=2.5in]{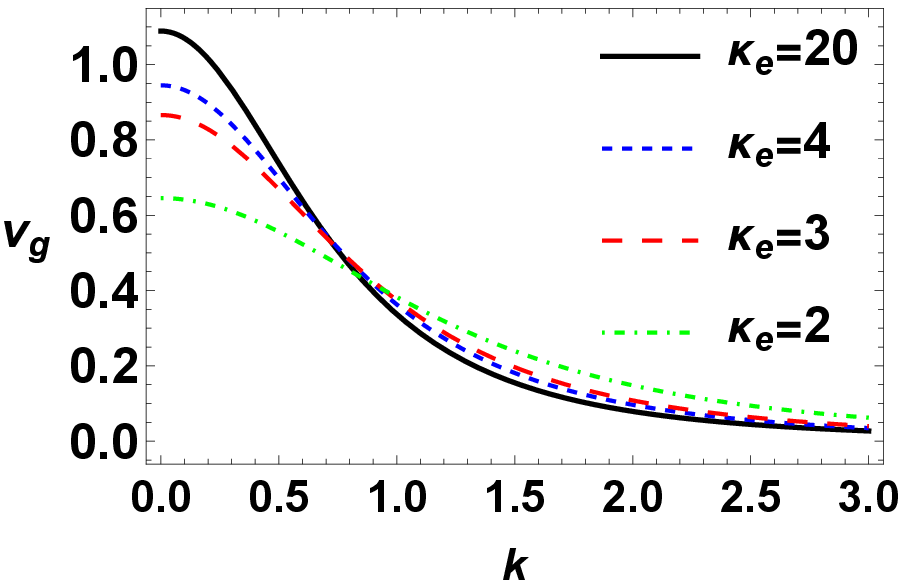}}
\caption{Plot of $v_g$ vs. $k$ for different values of (a) $\delta_d$  for fixed value of $\kappa_e=2$, (b) $\kappa_e$ for fixed value of $\delta_d=0.2$.}\label{f2}
\end{figure*}
\begin{figure*}
\centering
\subfigure[]{\includegraphics[width=2.5in]{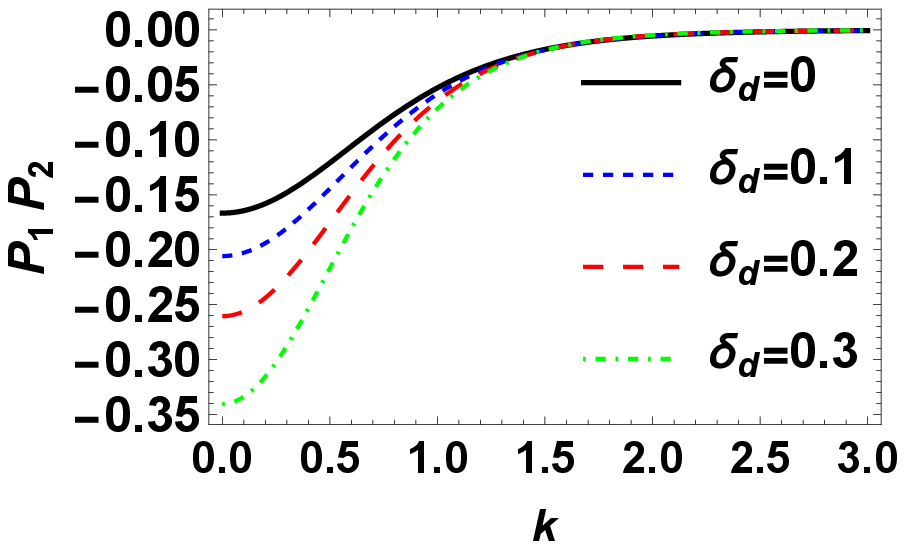}}
\subfigure[]{\includegraphics[width=2.5in]{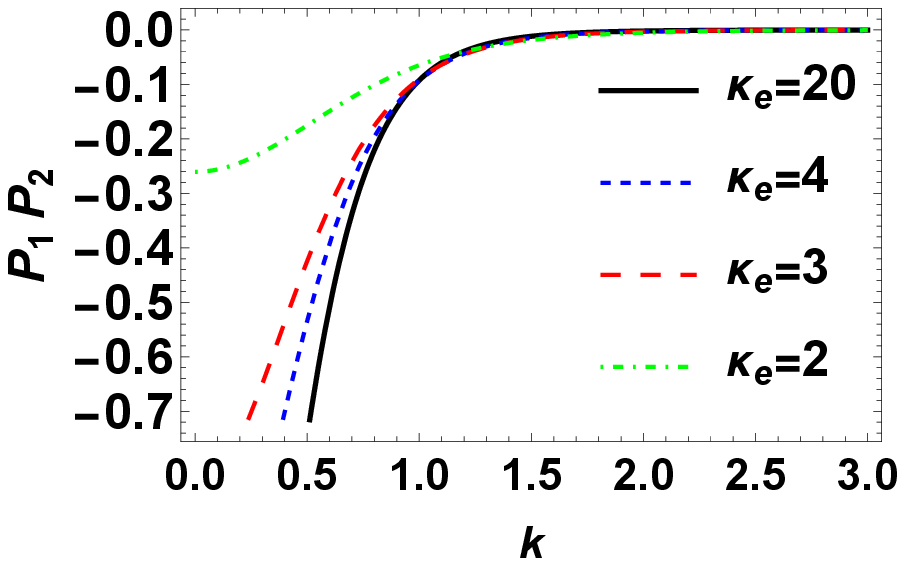}}
\caption{Plot of $P_1P_2$ vs. $k$ for different values of (a) $\delta_d$  for fixed value of $\kappa_e=2$, (b) $\kappa_e$ for fixed value of $\delta_d=0.2$.}\label{f3}
\end{figure*}
\begin{figure*}
\centering
\subfigure[]{\includegraphics[width=2.5in]{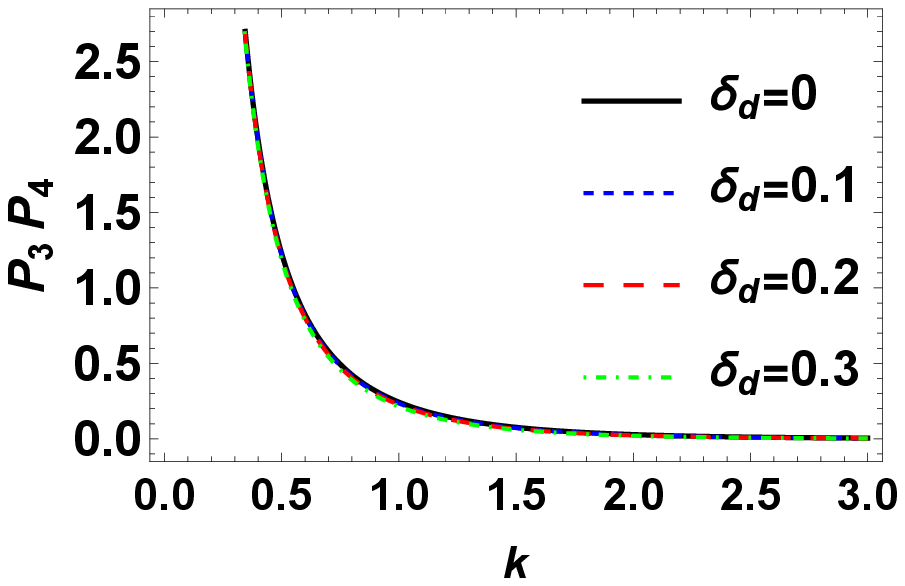}}
\subfigure[]{\includegraphics[width=2.5in]{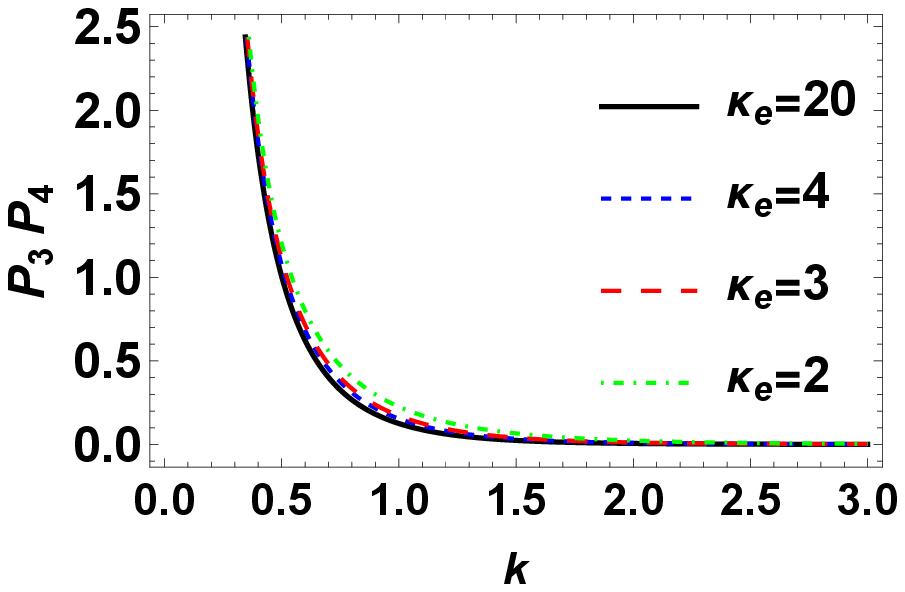}}
\caption{Plot of $P_3P_4$ vs. $k$ for different values of (a) $\delta_d$  for fixed value of $\kappa_e=2$, (b) $\kappa_e$ for fixed value of $\delta_d=0.2$.}\label{f4}
\end{figure*}

In 3$^{rd}$ order in $\eps$, the condition for annihilation of secular terms leads to a closed system of equations in the form:
\ba&& i\frac{\pd\psi}{\pd t}+P_1\frac{\pd^2\psi}{\pd x^2}+P_2\frac{\pd^2\psi}{\pd y^2}+Q_1\pss\psi+Q_2\psi Y=0\nn\\
&& P_3\frac{\pd^2 Y}{\pd x^2}+P_4\frac{\pd^2Y}{\pd y^2}+Q_3\frac{\pd^2\pss}{\pd y^2}=0  \label{dssystem} \ay
in terms of $\psi=\phi_{1}^{(1)}$ and $w=v_{y,z}^{(0)}$.
The independent variable appearing in the latter system of equations are actually $\{ x', y', t' \} = \{ X_1 - v_{g, x} T_1, Y_1 - v_{g, y} T_1, T_2\}$, but the primes have been  dropped for simplicity in the algebra to follow.

All coefficients in the Davey Stewartson equation (DS) system above are real and defined in the Appendix. Note that $P_1<0$ whereas $P_{2,3,4}>0$ for any values of the plasma parameters within the given (cold ions) fluid model.
\begin{figure*}
\centering
\subfigure[]{\includegraphics[width=2in,height=1.8in]{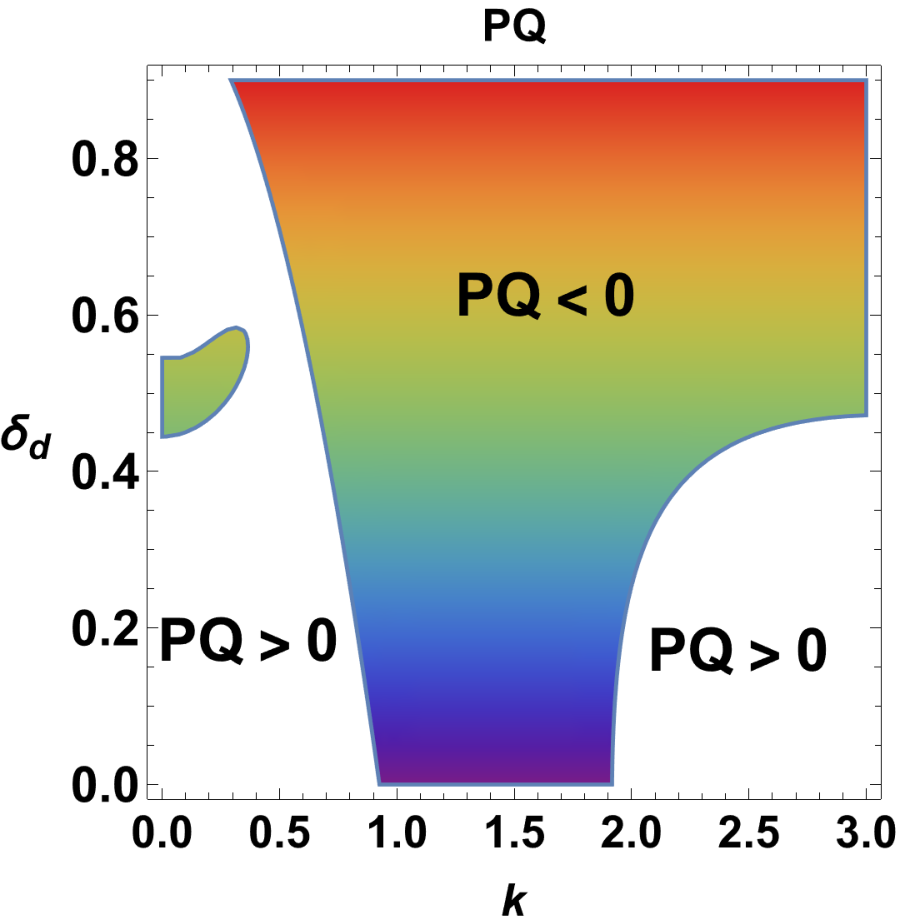}}
\subfigure[]{\includegraphics[width=2in,height=1.8in]{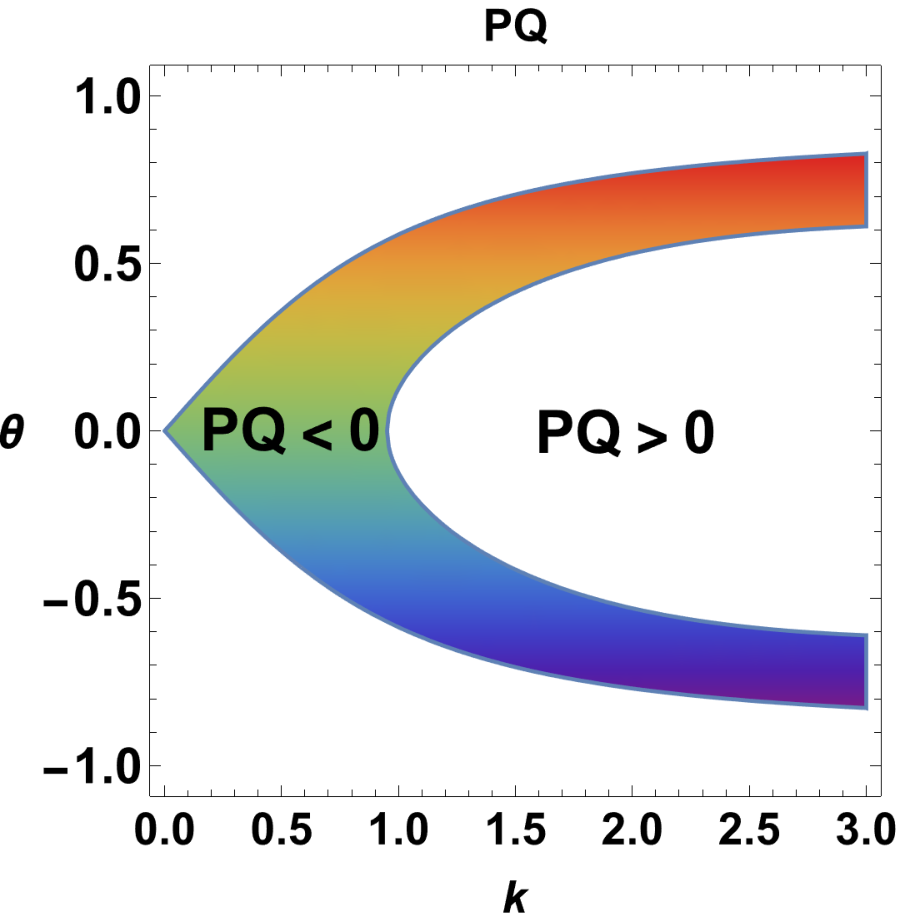}}
\subfigure[]{\includegraphics[width=2in,height=1.8in]{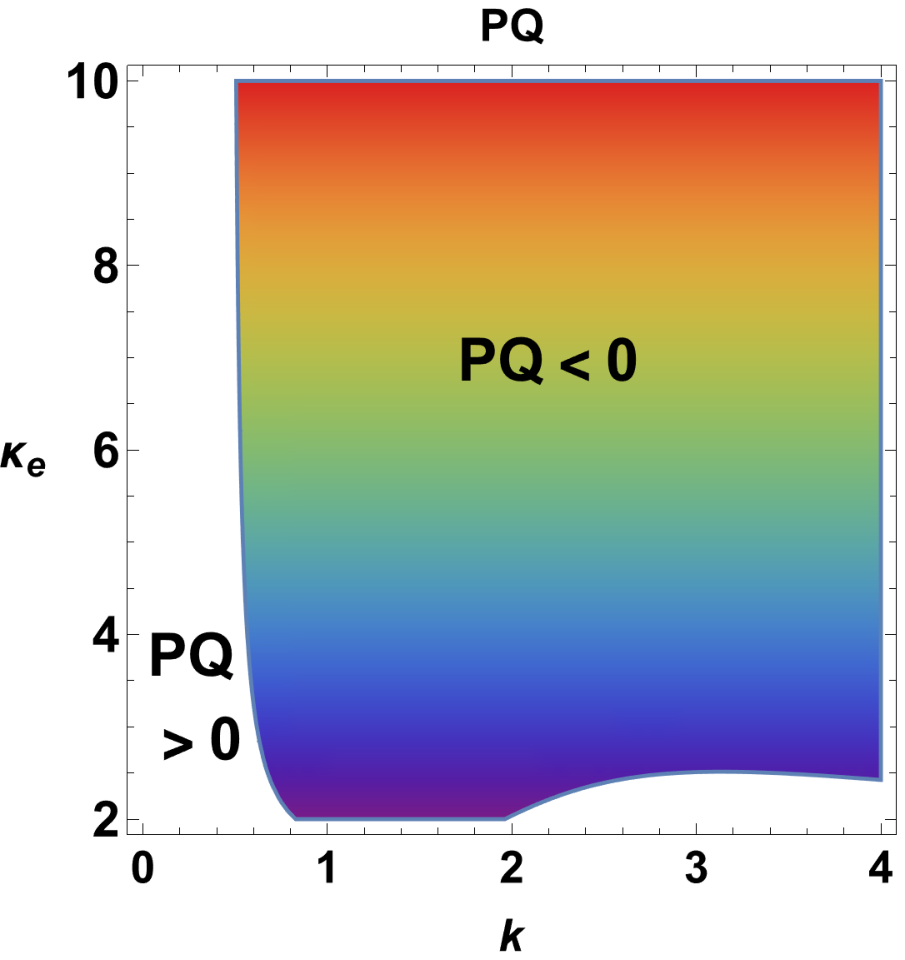}}
\caption{(Negative dust) Contour plot of $P Q$ in (a) $k$-$\delta_d$ plane for $\kappa_e=2$; (b) $k$-$\theta$ plane for $\kappa_e=2$ and $\delta_d=0.2$ (negative dust; $\delta_e = 0.8$); (c) $k$-$\kappa_e$ plane for $\delta_d=0.2$ (negative dust; $\delta_e = 0.8$).}  \label{f5}
\end{figure*}

\begin{figure*}
\centering
\subfigure[]{\includegraphics[width=2in,height=1.8in]{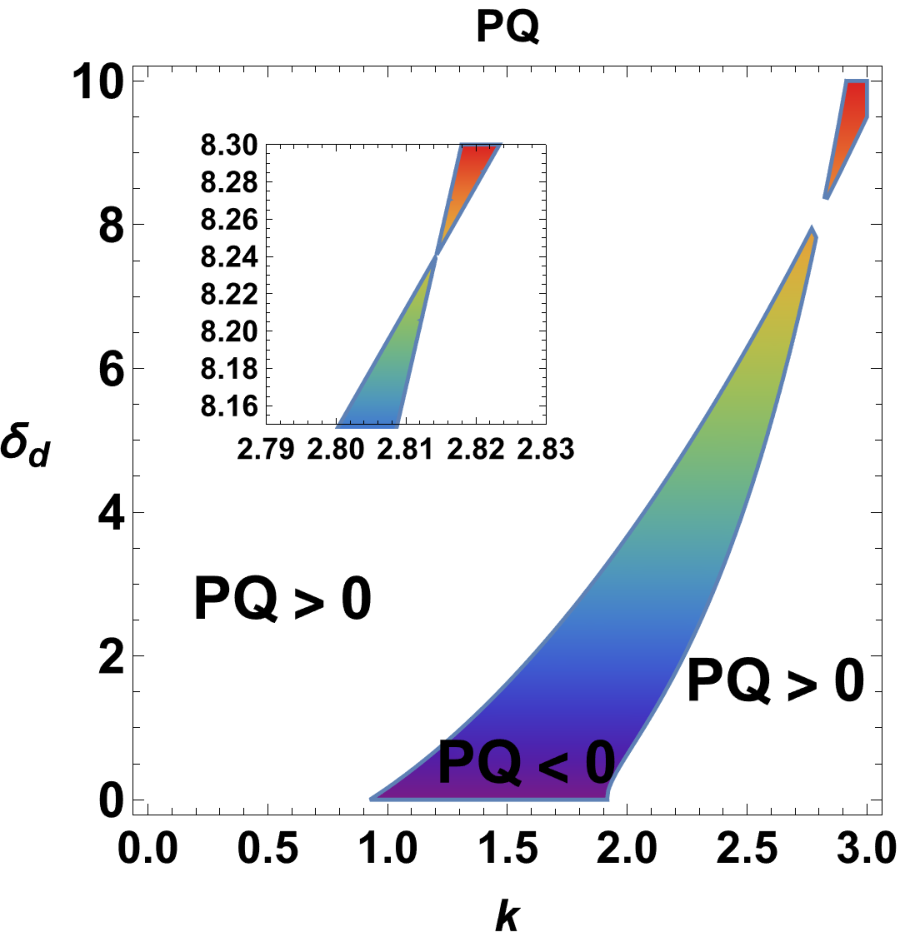}}
\subfigure[]{\includegraphics[width=2in,height=1.8in]{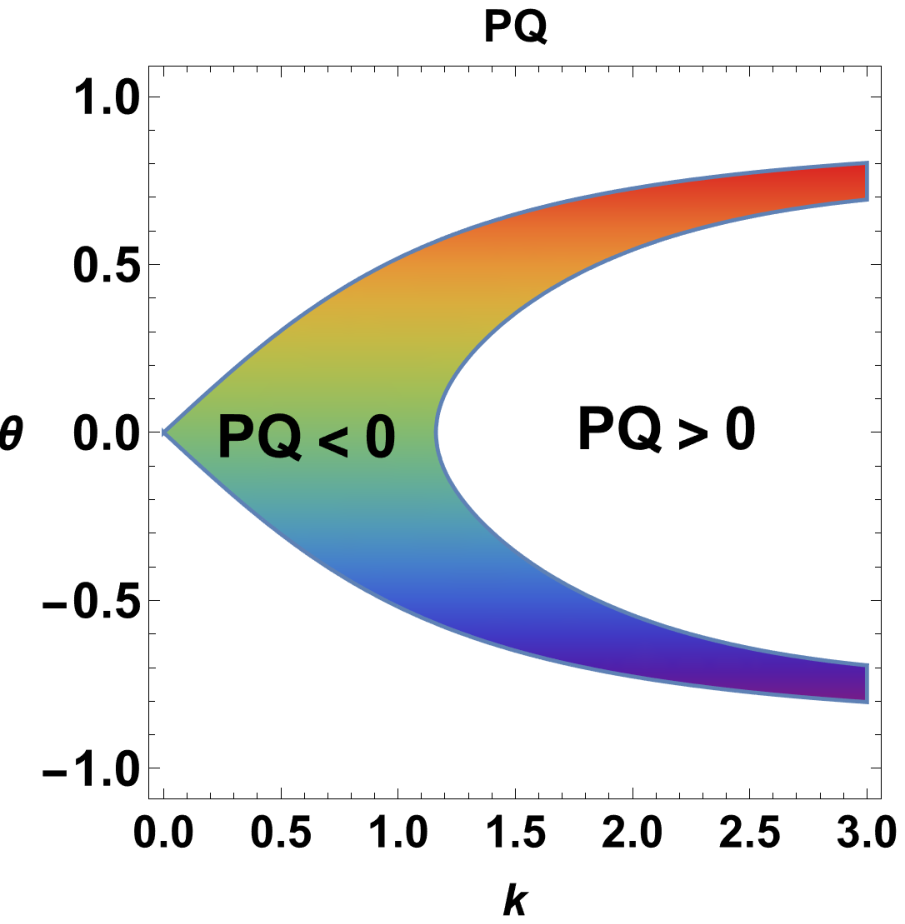}}
\subfigure[]{\includegraphics[width=2in,height=1.8in]{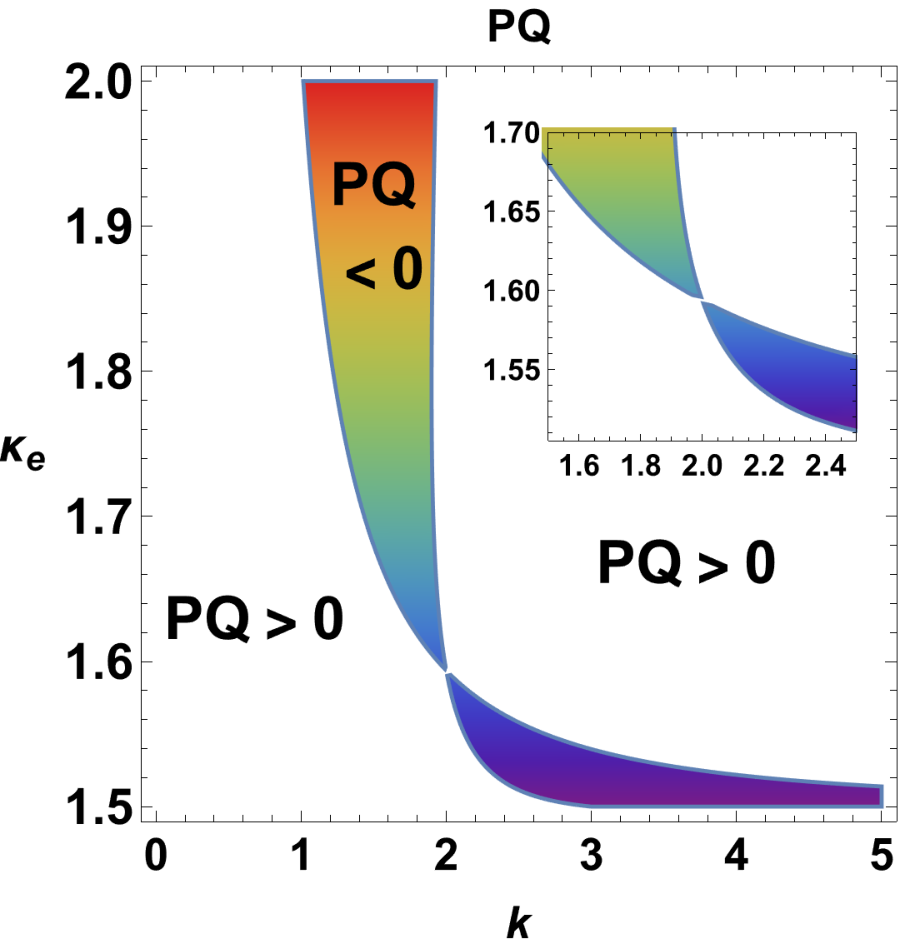}}
\caption{(Positive dust) Contour plot of $P Q$ in (a) $k$-$\delta_d$ plane for $\kappa_e=2$; (b) $k$-$\theta$ plane for $\kappa_e=2$ and $\delta_d = 0.2$ (positive dust; $\delta_e = 1.2$); (c) $k$-$\kappa_e$ plane for $\delta_d=0.2$ (positive dust; $\delta_e = 1.2$).}   \label{f5b}
\end{figure*}

\par
Given the sign(s) prescribed for the coefficients $P_{1, 2 , 3, 4}$ (also see Figs. (\ref{f3}) and (\ref{f4}) below), the DS system obtained in our case is hyperbolic-elliptic (i.e., of DS-II type), for any value of the model parameters. The hyperbolic-elliptic case occurs when $P_1P_2<0$ and $P_3P_4>0$. Dromions do not exist in this regime. Some prior investigations \citep{McConnell,Klein} have focused on numerical studies of singular, lump and rogue wave solutions. Rogue waves are found explicitly by Hirota's method in \citet{Ohta}. In \citet{Kavitha}, a solution is found in terms of exponentials for the general DS equation. The elliptic-hyperbolic case occurs when $P_1P_2>0$ and $P_3P_4<0$ and is commonly called the DS-I system. This was the first form found by Davey and Stewartson in their investigation into water waves \citep{ds}. For a range of values these equations can be solved by the inverse scattering method \citep{Fokas} and by Hirota's Bilinear method \citep{Satsuma}; in a nutshell, the following conclusions have been reached by those earlier studies: (i) for arbitrary time-independent boundary conditions, any arbitrary initial disturbance will decompose into a number of two dimensional breathers. Similarly, (ii) for arbitrary time-dependent boundary conditions, any arbitrary initial disturbance will decompose into a number of two-dimensional traveling localized structures. Since the two-dimensional localized solutions are associated with the discrete spectrum, it follows that they are nonlinear distortions of the bound states of the linearized equation. It turns out that in contrast to one-dimensional solitons these two-dimensional coherent solutions do not in general preserve their form upon interaction and exchange energy (only for a special choice of the spectral parameters these solutions preserve their form are revoked) \citep{Fokas}. Although non-trivial boundary conditions on $Y$ are required to obtain soliton or dromion solutions in this case \citep{White}.  When both (products) $P_1 P_2$ and $P_3 P_4$ are positive, the above DS system possesses solutions which take the form of a line soliton along the $x$-direction and are periodic in $y$ \citep{Groves}. This solution however will not occur in this particular model, given that the group velocity is a positive function with negative curvature. On the other hand,  when both $P_1 P_2$ and $P_3 P_4$ are negative, various types of solutions, in the form of rogue waves, breathers, solitons and hybrid versions can exist \citep{Rao}.

However, our model does not lead to the DS-I regime for any value of the parameters. This may presumably be the case if additional effects are taken into account, e.g. thermal ion pressure or an ambient magnetic field.

Earlier works have shown that the above system (DS-II) occurs in relation with ion acoustic \citep{Nishinari} and dust-ion acoustic \citep{xue04} plasmas. On the other hand, DS-I occurs in (un)magnetized dust acoustic \citep{aDuan,yashika} and electron acoustic \citep{Ghosh} plasmas, etc. Those systems sustain dromion solutions which, rather counter-intuitively, cannot exit in our model.

\section{Modulational Instability}

Let us investigate the stability of the DS system (\ref{dssystem}). A harmonic wave solution (equilibrium state) exists in the form $\psi_0=a_0e^{iQ_1a_0^2t}$ with
$Y=0$.
Assuming a harmonic variation (disturbance) off (but close to) that state, the equilibrium solution is modified as follows:
\ba\psi_0&=&a_0e^{iQ_1a_0^2t}\to a_0(1+a_1(x,y,t))e^{i(aQ_1^2t+b_1(x,y,t))}\nn\\
Y&=&0\to Y_1(x,y,t) \,  \nn \ay
where $a_1, b_1, Y_1$ are real functions.
 Separating real from imaginary parts and considering harmonic variations in the linearized system of the form:
\be\left(\begin{array}{c}a_1\\b_1\\ Y_1\end{array}\right)=\left(\begin{array}{c}\alpha\\\beta\\\gamma\end{array}\right)e^{i(kx+ly-\Omega t)}+c.c.,  \label{pert2}\e
we obtain a dispersion relation for the perturbation:
\begin{multline}
 \Omega^2(P_3k^2+P_4l^2)=(P_1k^2+P_2 l^2)\times\\
\left(2Q_2Q_3a_0^2l^2+(P_1k^2+P_2l^2-2Q_1a_0^2)(P_3k^2+P_4l^2)\right)\label{dis1}
\end{multline}
\begin{figure*}
\centering
\subfigure[]{\includegraphics[width=2.5in]{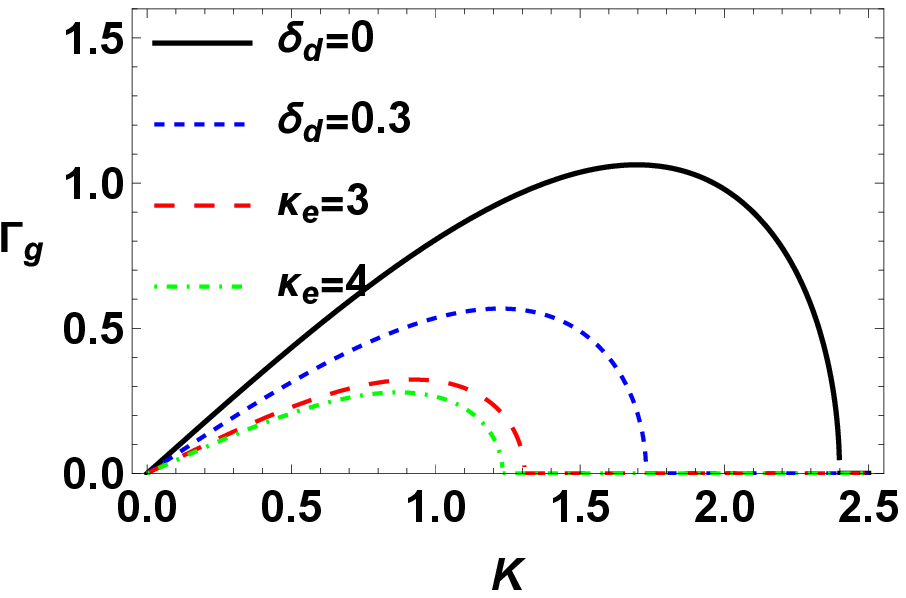}}
\subfigure[]{\includegraphics[width=2.5in]{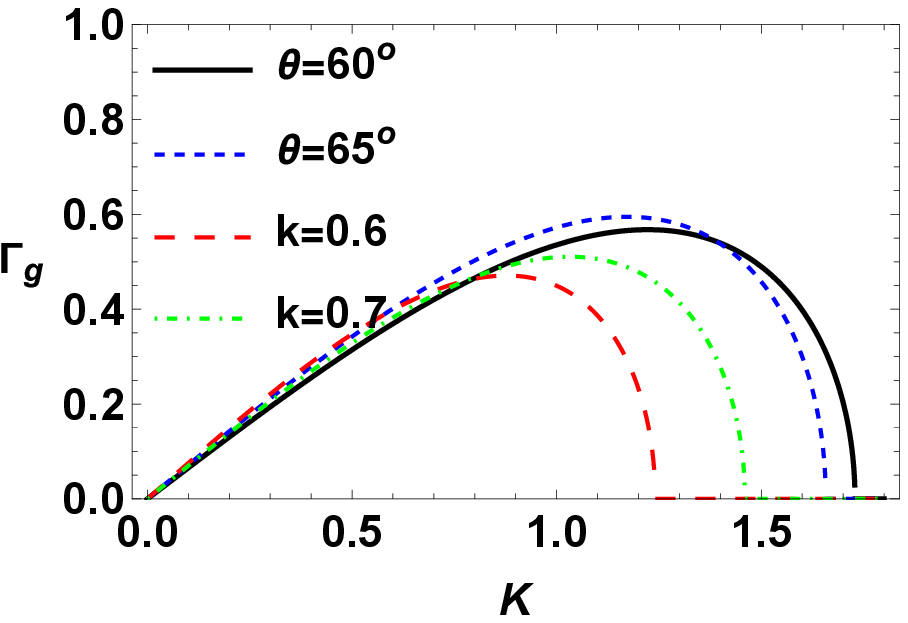}}
\caption{(Negative dust) Plot of the growth rate ($\Gamma_g$) versus the perturbation wavenumber ($K$), for different values of (a) $\delta_d$  and $\kappa_e$; (b) $\theta$ and fundamental wavenumber ($k$) for fixed for $\theta=55^{o}$, $k=2$, $\kappa_e=2$ and $\delta_d=0.2$ respectively.}\label{f6}
\end{figure*}
\begin{figure*}
\centering
\subfigure[]{\includegraphics[width=2.5in]{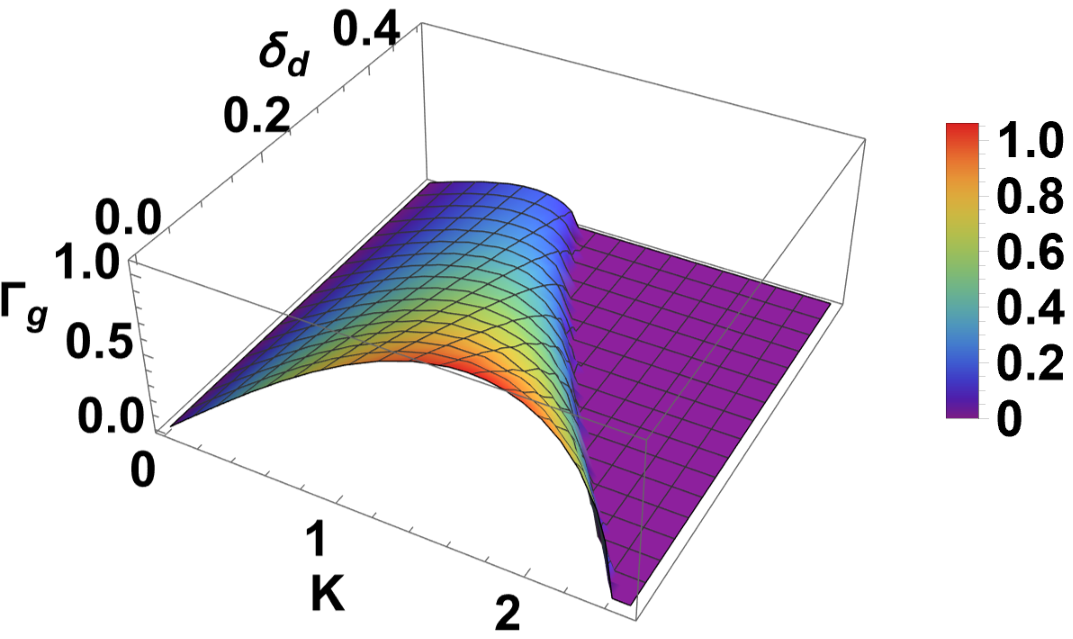}}
\subfigure[]{\includegraphics[width=2.5in]{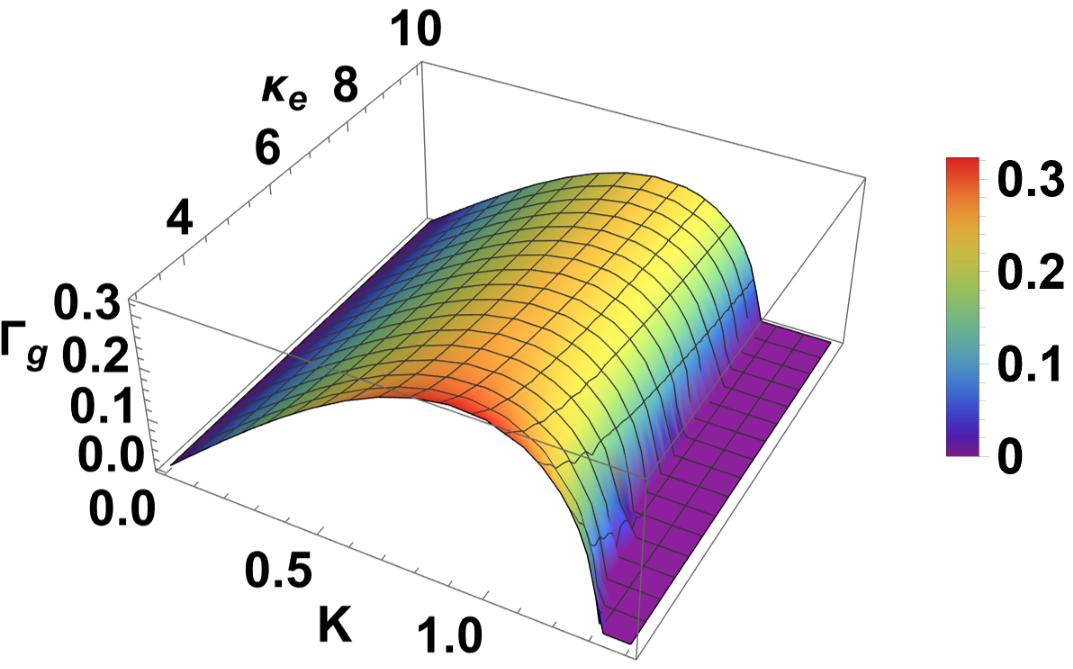}}
\caption{(Negative dust) 3D Plot of the growth rate ($\Gamma_g$) in the perturbation wavenumber $K$ -- $\delta_d$ plane (left panel) and in the $K$ - $\kappa_e$ plane (right panel) for fixed $\theta=55^{o}$, $k=2$, $\kappa_e=2$ and $\delta_d=0.2$ respectively.}\label{f6a}
\end{figure*}

\begin{figure*}
\centering
\subfigure[]{\includegraphics[width=2.5in]{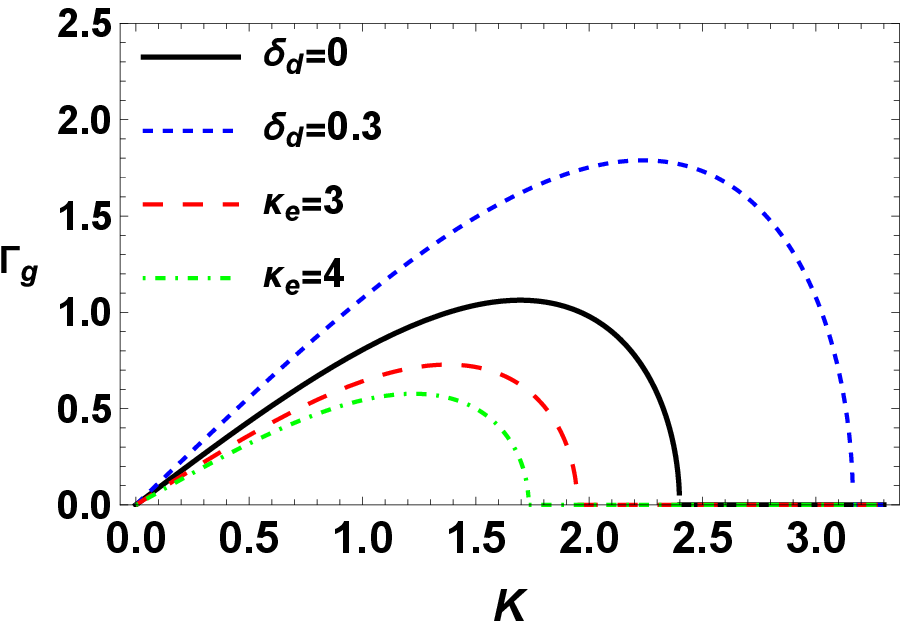}}
\subfigure[]{\includegraphics[width=2.5in]{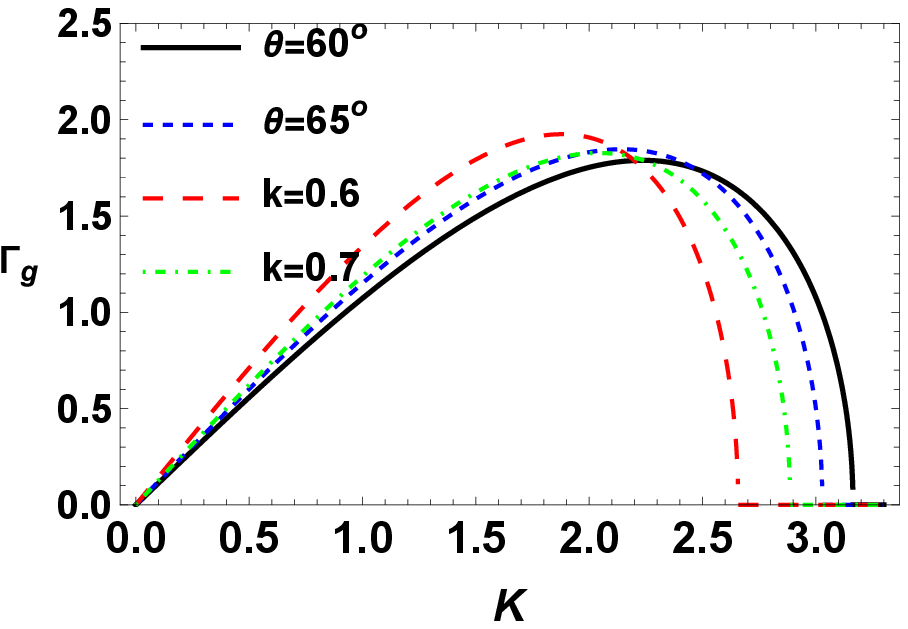}}
\caption{(Positive dust) lot of the growth rate ($\Gamma_g$) vs. the perturbation wavenumber ($K$) for different values of (a) $\delta_d$  and $\kappa_e$  (b) $\theta$ and fundamental wavenumber ($k$) for fixed for $\theta=55^{o}$, $k=2$, $\kappa_e=2$ and $\delta_d=0.2$ respectively.}\label{f6b}
\end{figure*}

\begin{figure*}
\centering
\subfigure[]{\includegraphics[width=2.5in]{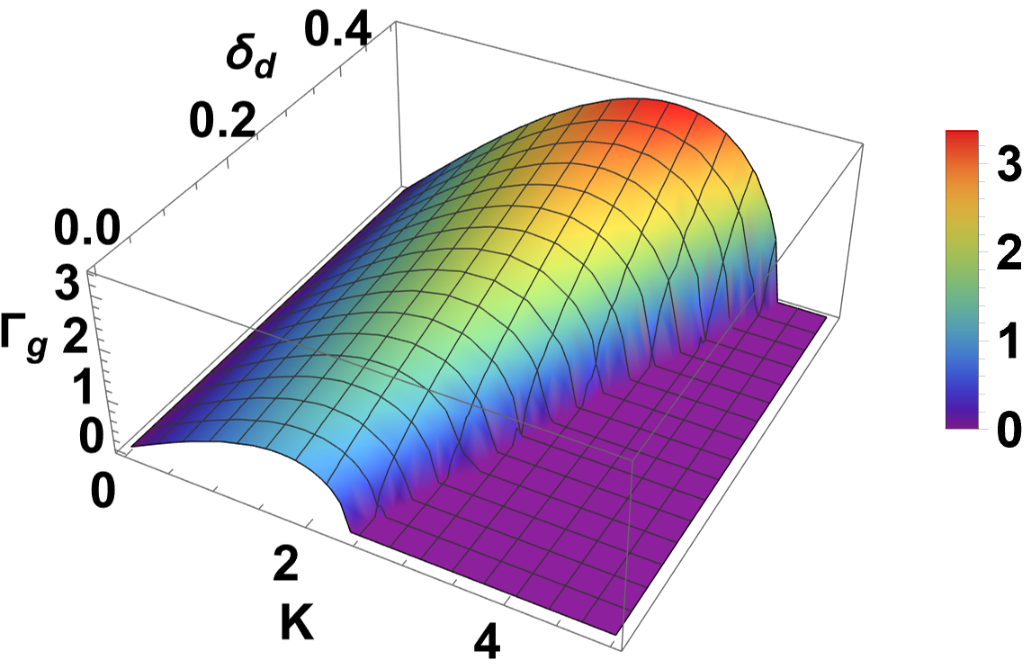}}
\subfigure[]{\includegraphics[width=2.5in]{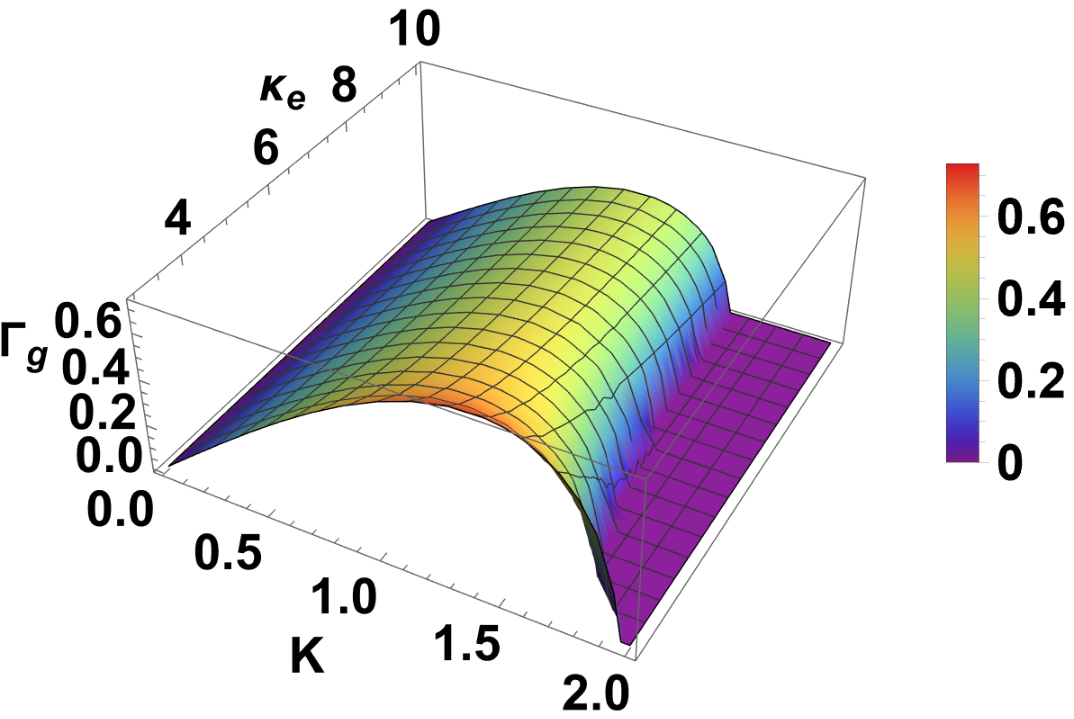}}
\caption{(Positive dust) 3D Plot of the growth rate ($\Gamma_g$) in the perturbation wavenumber $K$ - $\delta_d$ plane (b)  $K$ - $\kappa_e$ plane for fixed for $\theta=55^{o}$, $k=2$, $\kappa_e=2$ and $\delta_d=0.2$ respectively.}\label{f6c}
\end{figure*}
Provided that $P_3k^2+P_4l^2\ne0$, Eq. (\ref{dis1}) becomes
\begin{multline} \Omega^2=(P_1k^2+P_2l^2)^2\\
\left(1-\frac{2Q_1a_0^2}{P_1k^2+P_2l^2}-\frac{2Q_2Q_3a_0^2l^2}{(P_1k^2+P_2l^2)(P_3k^2+P_4l^2)}\right)\end{multline}
Furthermore, if $(k, l) = (k\cos \theta, k\sin \theta)$, this can be manipulated to obtain a form reminiscent of the 1D (NLS) case
\be \Omega^2= (PK^2)^2\left(1-\frac{2a_0^2Q}{PK^2}\right),\e
The expression of growth rate can be determined as
\be \Gamma_g=-Im(\Omega)= (PK^2)\left(\frac{2a_0^2Q}{PK^2}-1\right)^{1/2},\e

where
\ba P&=& P_1\cos^2\theta+P_2\sin^2\theta\nn\\
Q&=& Q_1+\frac{Q_2Q_3\sin^2\theta}{P_3\cos^2\theta+P_4\sin^2\theta} \ay

When $\theta=0$, the 1D dispersion relation $\Omega^2= (P_1K^2)^2\left(1-\frac{2a_0^2Q_1}{P_1K^2}\right)$ is recovered which is known to depict harmonic modulation of wavepackets in the NLS model \citep{ik05}.

\section{Parametric analysis}

In order to gain insight on the impact of various parameters on the dispersion characteristics of electrostatic waves, we have depicted in Fig. \ref{f1} the variation of $\omega$ vs. $k$ for different values of $\delta_d$  and $\kappa_e$. It is obvious that both the frequency and the phase speed of DIA wavepackets increase with higher  $\delta_d$ (i.e., for stronger dust concentration). On the contrary, lower values of  $\kappa_e$ (i.e., stronger deviation from the Maxwellian distribution) lead to a decrease in the frequency and in the phase velocity of DIA wavepacket. For clarity, see the 3D plots in Fig. \ref{f1b}.

Fig. \ref{f2} shows the variation of the group velocity for different values of $\delta_d$  and $\kappa_e$. The group velocity  increases with an increase in either $\delta_d$ (i.e., dust concentration) or $\kappa_e$ (i.e., decrease in the superthermality of electrons), in agreement with in Fig. \ref{f1}.

Fig. \ref{f3} shows that product $P_1P_2$ is negative whereas product $P_3P_4$ is positive (in Fig. \ref{f4}) for the given different values of $\delta_d$ and $\kappa_e$. Thus, our model leads to the DS-II regime for any value of the parameters.

Fig. \ref{f5} depicts a region plot of the product $P Q$  in various combination of the relevant parameters (the carrier wavenumber $k$, the dust density ratio $\delta_d$ and the spectral index $\kappa_e$).    
Recalling that $P Q < 0 $ is the criterion for stability, one realizes that the modulationally unstable region (where $P Q > 0 $) is represented in white color in the plots. The interface between stable and unstable regions in Fig. \ref{f5} represents the critical wave number $k (= k_c)$. For positive values of $P Q$, external perturbations make the envelope unstable, which may either lead to wavepacket collapse or presumably to the formation of  bright envelope structures (pulses). It is anticipated that the instability gets saturated by producing a train of envelope pulses (known as bright solitons in the 1D description). For $PQ < 0$, a stable wavepacket may propagate in the form of a  dark envelope (a localized envelope ``hole"), as known from the 1D case. The stability region (i.e., $PQ<0$) is between $k \approx $ 0.9 to 2 for the dust-free (electron-ion) case; however, as the negative dust concentration  increases, the region of stability  becomes wider in the wavenumber range of values (see Fig. \ref{f5}(a)).

In a similar way, the direction of the harmonic envelope perturbation considered above -- see (\ref{pert2})-(\ref{dis1}) (expressed via $\theta$), restricts the stability region for higher wavenumber whereas the unstable region expands (see Fig. \ref{f5}(b)). Fig. \ref{f5}(c)  illustrates that for lower values of $\kappa_e=(2,3)$, the stability region is narrow but for higher values it expands in the entire wavenumber range and the unstable region is only restricted between 0 to ($\approx$)0.6.  Similarly, Fig \ref{f5b} depicts a contour plot of the product $P Q$  in (a) $k$-$\delta_d$ plane (b) $k$-$\theta$ plane(c) $k$-$\kappa_e$ plane for positive dust.  The stability region (i.e., $PQ<0$) is between $k \approx $ 0.9 to 1.9 for the dust-free case, but as the positive dust concentration  increases, the stability  of DIA wavepackets is confined to a narrower wavenumber region in the advent of instability (see Fig. \ref{f5b} (a) wherein, for clarity, a zoom-in plot is embedded. Similarly, the angle of propagation of the envelope wave ($\theta$) restricts the stability region for higher wavenumber values, whereas the unstable region expands (see Fig. \ref{f5b} (b)). Fig. \ref{f5b}(c)  illustrates that the stability region shrinks in the wavenumber region, giving rise to full instability in that region.

\emph{Negative dust:} \ \ Fig. \ref{f6}(a) illustrates the variation of the growth rate of modulational instability for different values of  $\delta_d$  and $\kappa_e$, for negative dust. It is noted that for higher values of $\delta_d$ (i.e., for stronger dust concentration), the MI growth rate is suppressed. On the other hand, for smaller values of $\kappa_e$, i.e. for stronger deviation from the thermal (Maxwell-Boltzmann) equilibrium, the growth rate is enhanced. For more clarity see 3D Figs \ref{f6a}(a-b). Suprathermal electrons therefore lead to an increase in the modulational instability growth rate. Fig. \ref{f6}(b)illustrates the variation of the growth rate of modulational instability for different values of  $\theta$  and $k$, for negative dust. It is seen that the growth rate increases for higher values of  the fundamental wavenumber (k).  Furthermore, the growth rate decreases for higher values of the propagation angle ($\theta$).

\emph{Positive dust:} \ \ Fig \ref{f6b} (a) illustrates the variation of the MI growth rate of DIA wavepackets for different values of $\delta_d$  and $\kappa_e$, for positively charged dust. We see that for higher  $\delta_d$ (i.e., for stronger dust concentration) modulational instability is enhanced, and the same is true for smaller values of $\kappa_e$ (i.e. stronger deviation from the Maxwellian).  Fig. \ref{f6b}(b)illustrates the variation of the growth rate of modulational instability for different values of  $\theta$  and $k$, for negative dust. For more clarity see 3D Figs \ref{f6c} (a-b). It is seen that the growth rate increases for higher values of  the fundamental wavenumber (k).  Furthermore, the growth rate decreases for higher values of the propagation angle ($\theta$). Here, it is important to mention that negative or positive dust will modify the balance between the ion and the electron densities which leads to quantitative changes. For negative dust, a decrease in the density of electrons results in reduction in the modulation growth rate, because the nonlinear current is created by the motion of the electrons and decrease in the population of electrons leads to the decrease in the nonlinearity of the medium and consequently to a decrease in the growth rate. On the other hand, for positive dust grain charge, the same argument holds, in a reverse manner: for a fixed amount of dust (say, for a fixed dust-to-ion density ration, i.e. for a given value of the Havnes parameter, i.e. our $\delta_d$ parameter above), positive dust charge in combination with positive ions leads to a significant increase of the electron component (following a simple charge balance argument, as the electrons now have to balance the positive charge of both ions and dust), hence nonlinearity is expected to increase, again leading to an increased modulation growth rate.

\section{Conclusions}

We have analyzed a two-dimensional Davey–Stewartson (DS) equation for the evolution of modulated dust-ion acoustic wavepackets in non-Maxwellian dusty plasmas, taking into account the presence of a superthermal electron population and immobile dust in the background. The modulational (in)stability profile of DIA wavepackets for both positive as well as negative dust was investigated. A set of explicit criteria for modulational instability to occur was obtained. Stronger negative dust concentration (regadless of the values of $\kappa_e$) result in a narrower instability window in the $K$ (perturbation wavenumber) domain and to a suppressed growth rate. In the opposite trend, the modulational instability growth rate increases for positive dust concentration and the instability window gets larger - hence positive dust favors modulational instability. Finally, stronger deviation from Maxwell-Boltzmann equilbrium, i.e. smaller $\kappa_e$ values, lead(s) to stronger instability growth in a wider wavenumber window -- and this is true regardless of the dust charge sign (i.e. for either positive or negative dust). The  wavepacket modulation properties in 2D dusty plasmas thus differ from e.g. Maxwellian plasmas in 1D, both quanititatively and qualitatively, as indicated by a generalized dispersion relation (for the amplitude perturbation).
Our results are in agreement with the study by  \citet{xue04} (based on the 2D DS system) in the Maxwellian limit.  It is remarkable  that dimensionality alters the modulation behavior of DIA wavepackets significantly. Our results can be applied to existing experimental data in space, especially in Saturn's magnetosphere.

\section*{Acknowledgements}

The authors gratefully acknowledge financial support from Khalifa University of Science and Technology, Abu Dhabi UAE via the (internal funding) project FSU-2021-012/8474000352. Funding from the Abu Dhabi Department of Education and Knowledge (ADEK), currently ASPIRE UAE, via the AARE-2018 research grant ADEK/HE/157/18 is acknowledged. Author IK gratefully acknowledges financial support from Khalifa University’s Space and Planetary Science Center under grant No. KU-SPSC-8474000336.

%%%%%%%%%%%%%%%%%%%%%%%%%%%%%%%%%%%%%%%%%%%%%%%%%%

\section*{Data Availability}

The data underlying this article will be shared on reasonable request to the corresponding author.


%%%%%%%%%%%%%%%%%%%%%%%%%%%%% APPENDICES %%%%%%%%%%%%%%%%%%%%%
\appendix
\section{Coefficients in the DS system (22)}

The (real) coefficients in Eqs. (\ref{dssystem}) are given by:
\begin{multline} Q_1=\frac{\omega^3}{2k^2}\left(2c_2\left(\cad+\ccd\right)+3c_3\right)-k\left(k\ccc+\cac\right)-\frac{\omega}{2}\left(\cab+\ccb\right)=\nn\\
\cad\left(\frac{c_2\omega^3}{k^2}-\frac{k}{v_g}-\frac{c_1\omega}{2}\right)+
\ccd\left(\frac{c_2\omega^3}{k^2}-\frac{3k^2}{2\omega}\right)+\frac{3c_3\omega^3}{k^2}\nn \\-\frac{k^3}{\omega^2v_g}-c_2\omega-\frac{5k^4}{4\omega^3}\nn\\
 Q_2=\frac{\omega^3c_2}{k^2}\gamma_\phi-k\gamma_u-\frac{\omega}{2}\gamma_n=\frac{v_g}{c_1v_g^2-1}\left(\frac{\omega^3c_2}{k^2}-\frac{k}{v_g}-\frac{c_1\omega}{2}\right) \, , \nn\\
Q_3=-\cad    \, , \nn\\
P_1=\frac{1}{2}\frac{\partial^2 \omega}{\partial k_1^2}=-\frac{3c_1\omega^5}{2k^4}   \, , \qquad P_2 = \frac{1}{2}\frac{\partial^2 \omega}{\partial k_2^2}=\frac{c_1\omega^3}{k^4}   \, , \nn\\
P_3 = v_g   \, ,\qquad
P_4 = \frac{v_g}{1-c_1v_g^2}\label{coefs} \, . \nn \\  \end{multline}
Note that $P_1 < 0 < P_{2,3,4}>0$ within the given plasma model, for any values of the relevant parameters.

Of particular interest is the behaviour of these coefficients near $k\approx 0$ (long wavelength limit). The coefficients above can be approximated as:

\begin{multline}
Q_1\approx  \sqrt c_1\left(\frac{c_2}{c_1^2}-\frac{3}{2}\right)\left(\frac{c_2}{3}-\frac{c_1^2}{2}\right)\frac{1}{k}=\sqrt c_1\frac{\left(2c_2-3c_1^2\right)^2}{12c_1^2}\frac{1}{k}\nn\\
=\frac{\delta_{e}^{1/2}(\kappa_e-\frac{1}{2})^{1/2}[(\kappa_e+\frac{1}{2})^{2}+9\delta_{e}^{2}(\kappa_e-\frac{1}{2})^{2}-6\delta_{e}(\kappa_e-\frac{1}{2})(\kappa_e+\frac{1}{2})]}{12(\kappa_e-\frac{3}{2})^{5/2}}\frac{1}{k} \, ,  \nn \\
Q_2 \approx -\frac{c_1}{3}\left(\frac{1}{c_1^2}-\frac{3}{2}\right)\frac{1}{k}=-\left(\frac{(\kappa_e-\frac{3}{2})}{3\delta_e(\kappa_e-\frac{1}{2})}+\frac{\delta_e(\kappa_e-\frac{1}{2})}{(\kappa_e-\frac{3}{2})}\right)\frac{1}{k}\nn\\
Q_3 \approx\frac{1}{3k^2}\left(3c_1^2-2c_2\right)= \frac{\delta_e(\kappa_e-\frac{1}{2})}{3k^2}\left(\frac{3\delta_e(\kappa_e-\frac{1}{2})-(\kappa_e+\frac{1}{2})}{(\kappa_e-\frac{3}{2})^2}\right) \, , \nn\\
P_1 \approx -\frac{3}{2c_1}\sqrt{\frac{1}{c_1}}k=-\frac{3k}{2}\left(\frac{\kappa_e-\frac{3}{2}}{\delta_e(\kappa_e-\frac{1}{2})}\right)^{3/2} \, , \nn\\
P_2 \approx \sqrt{\frac{1}{c_1}}\frac{1}{2k}=\sqrt{\frac{\kappa_e-\frac{3}{2}}{\delta_e(\kappa_e-\frac{1}{2})}}\frac{1}{2k} \, , \nn\\
P_3 \approx  \sqrt{\frac{1}{c_1}}\left(1-\frac{3k^2}{2c_1}\right)=\sqrt{\frac{\kappa_e-\frac{3}{2}}{\delta_e(\kappa_e-\frac{1}{2})}}
-\frac{3}{\delta_e^{3/2}} k^2 \left(\frac{\kappa_e-\frac{3}{2}}{\kappa_e-\frac{1}{2}}\right)^{3/2} \, , \nn\\
P_4 \approx \frac{1}{3}\sqrt{c_1}\frac{1}{k^2}= \frac{1}{3}\sqrt{\frac{\delta_e(\kappa_e-\frac{1}{2})}{(\kappa_e-\frac{3}{2})}}\frac{1}{k^2}\, .\nn\\
\end{multline}

%%%%%%%%%%%%%%%%%%%%%%%%%%%%%%%%%%%%%%%%%%%%%%%%%%

% Don't change these lines
\bsp	% typesetting comment
\label{lastpage}
\end{document}